\documentclass[12pt,a4paper]{conference}

\usepackage{fancyhdr}
\usepackage{graphicx,amsmath,amssymb,cite}
\usepackage{multind}
\makeindex{author} \makeindex{subject}
\newcommand{\beq}{\begin{equation}}
\newcommand{\eeq}[1]{\label{#1}\end{equation}}
\newcommand{\eeqn}{\end{equation}}

\newcommand{\beqa}{\begin{eqnarray}}
\newcommand{\eeqa}[1]{\label{#1}\end{eqnarray}}
\newcommand{\eeqan}{\end{eqnarray}}

\let\bar=\overbar

\newcommand{\Dslash}{\not{\hbox{\kern-4pt $D$}}}
\newcommand{\dslash}{\not{\hbox{\kern-2pt $\del$}}}

\newcommand{\msb}{{\bar{\ssstyle M \kern -1pt S}}}

\newcommand{\gev}{\ \text{GeV}}
\newcommand{\w}{\omega}

\newcommand{\UChPT}{UChPT}
\newcommand{\ChPT}{ChPT}

\begin{document}
\Chapter{S-wave meson scattering up to $\sqrt{s} \lesssim 2 \gev$ from chiral Lagrangians}
	   {S-wave meson scattering in UChPT}{M. Albaladejo \it{et al.}}
\vspace{-6 cm}\includegraphics[width=6 cm]{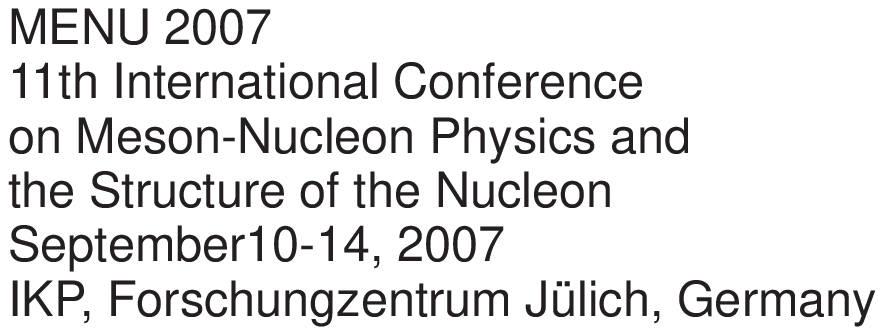}
\vspace{4 cm}

\addcontentsline{toc}{chapter}{{\it M. Albaladejo}} \label{authorStart}

\begin{raggedright}

M. Albaladejo$^{\star}$$^,$\footnote{E-mail address: \texttt{albaladejo@um.es}}\index{author}{Albaladejo, M.}, J.A. Oller$^{\star}$$^,$\footnote{E-mail address: \texttt{oller@um.es}}\index{author}{Oller, J.A.}, C. Piqueras$^{\dagger}\index{author}{Piqueras, C.}$
\bigskip\bigskip

$^{\star}$ Departamento de F\'isica, Universidad de Murcia, E-30071, Murcia, Spain\\
$^{\dagger}$ I.E.S. Ribera de los Molinos, E-30170 Murcia, Spain

\end{raggedright}

\begin{center}
\textbf{Abstract}
\end{center}

The problem of scalar mesons still remains a challenging puzzle, for which we do not even know which are the right pieces to set up. The proliferation of resonances (some of them are very broad and appear on top of hadronic thresholds) and of coupled channels that interact strongly among each other makes the study of this sector a hard task. Our objective is the study of the strongly interacting mesons in coupled channels with quantum numbers $J^{PC} = 0^{++}$ and $I=0$ and $I=1/2$, up to a center of mass energy $\sqrt{s} \lesssim 2 \gev$. Our framework is based on Unitary Chiral Perturbation Theory. We include for $I=0$ the channels: $\pi\pi$, $K\overline{K}$, $\eta\eta$, $\sigma\sigma$, $\eta\eta'$, $\rho\rho$, $\omega\omega$, $\eta'\eta'$, $\omega\phi$, $\phi\phi$,  $K^\ast \overline{K}^\ast$, $a_1(1260)\pi$ and $\pi^{\star}(1300)\pi$. In addition, and in order to constrain our fits, we also study the $I=1/2$, $3/2$ channels given by $K\pi$, $K\eta$ and $K\eta'$. We finally present  the resonant content of our fits with the $\sigma$, $f_0(980)$, $f_0(1310)$, $f_(1500)$, $f_0(1710)$ and $f_0(1790)$.

\section{Lagrangians. $U(3)$ symmetry}
Due to the spontaneous breakdown of chiral $SU(3)$ symmetry the 
$\pi$, $K$ and $\eta$ are the octet of pseudo-Goldstone bosons. As it is well known, chiral symmetry strongly constrains 
the allowed interactions between these pseudoscalars and it is a basic ingredient in any study of strong interactions involving
those mesons. 
If one considers higher energy regions, as it is our case here where we study the $I=0$ and $1/2$ S-waves up to 
about 2 GeV, 
one also needs to take into account the $\eta\eta$, $\eta\eta'$ and $\eta'\eta'$ channels. 
 Interestingly, in the large $N_c$ limit, the $\eta_1$ becomes the ninth 
 Goldstone boson. This fact can be used to settle down chiral Lagrangians based on $U(3)$ chiral symmetry 
 and to include the  $\eta_1$ field. The  $\eta_1$-$\eta_8$ mixing angle is taken as  $\sin \theta = -1/3 \to \theta\approx -20^\circ$.
 The channels we include  for $I=0$ are: (1) $\pi\pi$, (2) $K\overline{K}$, (3) $\eta\eta$, (4) $\sigma \sigma$, 
  (5) $\eta\eta'$, (6) $\rho\rho$, (7) $\w\w$, (8) $\eta\eta'$, (9) $\w\phi$, (10) $\phi\phi$, 
 (11) $K^\ast \overline{K}^\ast$, (12) $a_1(1260)\pi$ and (13) $\pi^{\star}(1300)\pi$.
 For $I=1/2$ and 3/2 we take the $K\pi$, $K\eta$ and $K\eta'$ ones. For these latter isospin channels 
  we follow ref.\cite{i-one-half} .

We employ the Chiral Perturbation Theory  (ChPT)  Lagrangians to lowest order, 
and the chirally invariant resonance ones \cite{lagr-ref}, with 
the $J^{PC} = 0^{++}$ singlet and octet multiplets. These Lagrangians also incorporate  
the $r_\mu$ and $l_\mu$ external sources by gauging the $U(3)_L\otimes U(3)_R$ chiral symmetry. In our case, 
as we are interested in the vector resonances, we have $r_{\mu} = l_{\mu} = g v_{\mu}$,
where the last term is a matrix of vector fields times a coupling constant, which can be determined through the width of $\rho \rightarrow \pi\pi$, $g=4.23$.

\section{Unitarization. $\sigma\sigma$ channels. Width effects}
We want to calculate the amplitudes involving the $\sigma\sigma$ channel from our Lagrangians. As it was shown within 
\UChPT\index{subject}{Chiral Perturbation Theory} \cite{sigma-ref} \index{subject}{sigma resonance}, the
 $\sigma$ is made up from two pions interacting in $I=0$ S-wave, which allows us to obtain these amplitudes without any new free parameter. 
 We first consider  the amplitudes of a generic channel, say $a$, to four pions grouped as two $I=0$ $\pi\pi$ states ($(\pi\pi)_0$). Let us call $s_1$ 
 and $s_2$ the total CM energy squared of each of these states. Now, every $(\pi\pi)_0$ state rescatters and gives
  rise to a $\sigma$ pole; 
 this is taken into account with the factor $1/D(s_1)D(s_2)$, where $D(s) = (1+t_2 G(s))^{-1}$, 
 $t_2$ is the elastic $(\pi\pi)_0$ S-wave amplitude at lowest order in \ChPT, and $G(s)$ is the two-meson loop function.
  The transition amplitude is obtained by taking the limits $s_i \rightarrow s_\sigma$, with $s_\sigma$ the $\sigma$ pole position. 
  If we denote the $a \rightarrow (\pi\pi)_0 (\pi\pi)_0$ amplitude by $T_{a\rightarrow(\pi\pi)_0(\pi\pi)_0}$,
   and the one deduced from \ChPT\ with resonances by $T_{a\rightarrow(\pi\pi)_0(\pi\pi)_0}^{2+R}$,  we have:
\begin{equation}\label{eq1}
T_{a\rightarrow(\pi\pi)_0(\pi\pi)_0} = T_{a\rightarrow(\pi\pi)_0(\pi\pi)_0}^{2+R} \frac{1}{D(s_1)D(s_2)} \text{.}
\end{equation}
 The $a \rightarrow (\sigma\sigma)_0$ amplitude, $N_{a\rightarrow (\sigma\sigma)_0}$, is obtained from,
\begin{equation}\label{eq2}
\lim_{s_i \rightarrow s_\sigma} T_{a\rightarrow(\pi\pi)_0(\pi\pi)_0} = \lim_{s_i \rightarrow s_\sigma} \frac{T_{a\rightarrow(\pi\pi)_0(\pi\pi)_0}^{2+R} }{D_{II}(s_1)D_{II}(s_2)}= N_{a\rightarrow (\sigma\sigma)_0} \frac{g^2_{\sigma\pi\pi}}{(s_1-s_\sigma)(s_2 - s_\sigma)}
\end{equation}
The subscript II means that we have to calculate the corresponding
function on the second Riemann sheet, where the $\sigma$ pole appears.
 Finally, calculating this limit with an appropriate Laurent expansion around $s_\sigma$, 
\begin{equation}
N_{a\rightarrow (\sigma\sigma)_0} = T_{a \rightarrow (\pi\pi)_0(\pi\pi)_0}^{2+R}
\left( \frac{\alpha_0}{g_{\sigma\pi\pi}} \right)^2 \quad \text{,}\quad \left( \frac{\alpha_0}{g_{\sigma\pi\pi}} \right)^2 \simeq 9.1\cdot
10^{-3} \gev^2~.
\end{equation}

Now, the general expression for a coupled channels partial wave amplitude is $T = (I+N(s)g(s))^{-1} N(s)$, where $N(s)$  is a matrix containing our amplitudes between all the channels.
  Each element of the diagonal matrix $g(s)$ is given by the once subtracted dispersion relation,
\begin{equation}\label{loopfunc}
g_i(s) = g_i(s_0) - \frac{s-s_0}{8\pi^2} \int_{s_{\text{th,i}}}^{\infty}\text{d}s' \frac{{p_i(s')}/{\sqrt{s'}}}{(s'-s_0)(s'-s+i\epsilon)}
\end{equation}
A remark is in order. These integrals involve the masses of the particles of the scattering states, but some of them, 
as the $\sigma$, $\rho$, $a_1(1260)$ and $\pi^{\star}(1300)$ have very large widths. To take these effects into account, we consider instead of 
 eq.\eqref{loopfunc}
an integral of this loop function times a mass  distribution over a wide range of masses for each of these unstable particles.

\section{Results and spectroscopy}
\begin{figure}[h]
\begingroup%
  \makeatletter%
  \newcommand{\GNUPLOTspecial}{%
    \@sanitize\catcode`\%=14\relax\special}%
  \setlength{\unitlength}{0.1bp}%
\begin{picture}(4320,3240)(0,0)%
\special{psfile=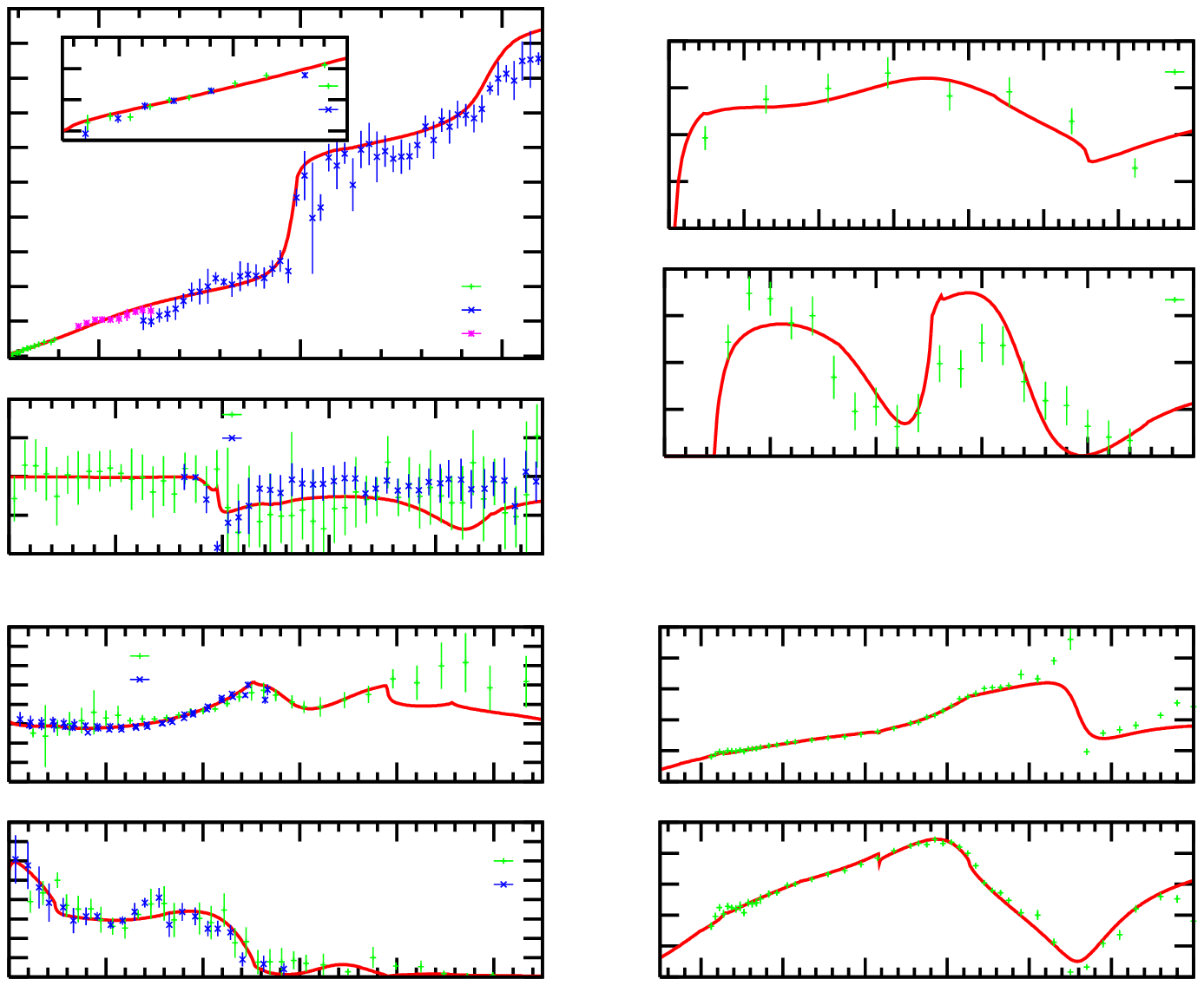 llx=0 lly=0 urx=432 ury=324 rwi=4320}
\fontsize{6}{\baselineskip}\selectfont
\put(3986,3108){\makebox(0,0)[r]{Binon et al.}}%
\put(3225,2499){\makebox(0,0){$\sqrt{s}\ \textrm{(MeV)}$}}%
\put(2190,2899){%
\special{ps: gsave currentpoint currentpoint translate
270 rotate neg exch neg exch translate}%
\makebox(0,0)[b]{\shortstack{$\lvert S_15 \rvert^2$}}%
\special{ps: currentpoint grestore moveto}%
}%
\put(4095,2559){\makebox(0,0){ 1850}}%
\put(3846,2559){\makebox(0,0){ 1800}}%
\put(3598,2559){\makebox(0,0){ 1750}}%
\put(3349,2559){\makebox(0,0){ 1700}}%
\put(3101,2559){\makebox(0,0){ 1650}}%
\put(2852,2559){\makebox(0,0){ 1600}}%
\put(2604,2559){\makebox(0,0){ 1550}}%
\put(2355,2559){\makebox(0,0){ 1500}}%
\put(2340,3210){\makebox(0,0)[r]{ 0.1}}%
\put(2340,3055){\makebox(0,0)[r]{ 0.075}}%
\put(2340,2900){\makebox(0,0)[r]{ 0.05}}%
\put(2340,2744){\makebox(0,0)[r]{ 0.025}}%
\put(2340,2589){\makebox(0,0)[r]{ 0}}%
\put(3986,2352){\makebox(0,0)[r]{Binon et al.}}%
\put(3217,1743){\makebox(0,0){$\sqrt{s}\ \textrm{(MeV)}$}}%
\put(2190,2143){%
\special{ps: gsave currentpoint currentpoint translate
270 rotate neg exch neg exch translate}%
\makebox(0,0)[b]{\shortstack{$\lvert S_13 \rvert^2$}}%
\special{ps: currentpoint grestore moveto}%
}%
\put(4095,1803){\makebox(0,0){ 2000}}%
\put(3744,1803){\makebox(0,0){ 1800}}%
\put(3393,1803){\makebox(0,0){ 1600}}%
\put(3042,1803){\makebox(0,0){ 1400}}%
\put(2691,1803){\makebox(0,0){ 1200}}%
\put(2340,1803){\makebox(0,0){ 1000}}%
\put(2325,2454){\makebox(0,0)[r]{ 0.2}}%
\put(2325,2299){\makebox(0,0)[r]{ 0.15}}%
\put(2325,2144){\makebox(0,0)[r]{ 0.1}}%
\put(2325,1988){\makebox(0,0)[r]{ 0.05}}%
\put(2325,1833){\makebox(0,0)[r]{ 0}}%
\put(3210,663){\makebox(0,0){$\sqrt{s}\ \textrm{(MeV)}$}}%
\put(2190,1009){%
\special{ps: gsave currentpoint currentpoint translate
270 rotate neg exch neg exch translate}%
\makebox(0,0)[b]{\shortstack{$\phi\ \textrm{(I=1/2)}$}}%
\special{ps: currentpoint grestore moveto}%
}%
\put(4095,723){\makebox(0,0){ 2000}}%
\put(3823,723){\makebox(0,0){ 1800}}%
\put(3550,723){\makebox(0,0){ 1600}}%
\put(3278,723){\makebox(0,0){ 1400}}%
\put(3006,723){\makebox(0,0){ 1200}}%
\put(2733,723){\makebox(0,0){ 1000}}%
\put(2461,723){\makebox(0,0){ 800}}%
\put(2310,1266){\makebox(0,0)[r]{ 250}}%
\put(2310,1163){\makebox(0,0)[r]{ 200}}%
\put(2310,1061){\makebox(0,0)[r]{ 150}}%
\put(2310,958){\makebox(0,0)[r]{ 100}}%
\put(2310,856){\makebox(0,0)[r]{ 50}}%
\put(2310,753){\makebox(0,0)[r]{ 0}}%
\put(3210,15){\makebox(0,0){$\sqrt{s}\ \textrm{(MeV)}$}}%
\put(2190,361){%
\special{ps: gsave currentpoint currentpoint translate
270 rotate neg exch neg exch translate}%
\makebox(0,0)[b]{\shortstack{$\lvert S \rvert\ \textrm{(I=1/2)}$}}%
\special{ps: currentpoint grestore moveto}%
}%
\put(4095,75){\makebox(0,0){ 2000}}%
\put(3823,75){\makebox(0,0){ 1800}}%
\put(3550,75){\makebox(0,0){ 1600}}%
\put(3278,75){\makebox(0,0){ 1400}}%
\put(3006,75){\makebox(0,0){ 1200}}%
\put(2733,75){\makebox(0,0){ 1000}}%
\put(2461,75){\makebox(0,0){ 800}}%
\put(2310,618){\makebox(0,0)[r]{ 250}}%
\put(2310,515){\makebox(0,0)[r]{ 200}}%
\put(2310,413){\makebox(0,0)[r]{ 150}}%
\put(2310,310){\makebox(0,0)[r]{ 100}}%
\put(2310,208){\makebox(0,0)[r]{ 50}}%
\put(2310,105){\makebox(0,0)[r]{ 0}}%
\put(1759,412){\makebox(0,0)[r]{Cohen et al.}}%
\put(1759,490){\makebox(0,0)[r]{Etkin et al.}}%
\put(1050,15){\makebox(0,0){$\sqrt{s}\ \textrm{(MeV)}$}}%
\put(30,361){%
\special{ps: gsave currentpoint currentpoint translate
270 rotate neg exch neg exch translate}%
\makebox(0,0)[b]{\shortstack{$\lvert S_{12} \rvert^2$}}%
\special{ps: currentpoint grestore moveto}%
}%
\put(1774,75){\makebox(0,0){ 2000}}%
\put(1452,75){\makebox(0,0){ 1800}}%
\put(1130,75){\makebox(0,0){ 1600}}%
\put(809,75){\makebox(0,0){ 1400}}%
\put(487,75){\makebox(0,0){ 1200}}%
\put(165,75){\makebox(0,0){ 1000}}%
\put(150,618){\makebox(0,0)[r]{ 0.8}}%
\put(150,554){\makebox(0,0)[r]{ 0.7}}%
\put(150,490){\makebox(0,0)[r]{ 0.6}}%
\put(150,426){\makebox(0,0)[r]{ 0.5}}%
\put(150,362){\makebox(0,0)[r]{ 0.4}}%
\put(150,297){\makebox(0,0)[r]{ 0.3}}%
\put(150,233){\makebox(0,0)[r]{ 0.2}}%
\put(150,169){\makebox(0,0)[r]{ 0.1}}%
\put(150,105){\makebox(0,0)[r]{ 0}}%
\put(552,1092){\makebox(0,0)[r]{Cohen et al.}}%
\put(552,1170){\makebox(0,0)[r]{Etkin et al.}}%
\put(1050,663){\makebox(0,0){$\sqrt{s}\ \textrm{(MeV)}$}}%
\put(30,1009){%
\special{ps: gsave currentpoint currentpoint translate
270 rotate neg exch neg exch translate}%
\makebox(0,0)[b]{\shortstack{$\delta_{12}\ \textrm{(deg)}$}}%
\special{ps: currentpoint grestore moveto}%
}%
\put(1774,723){\makebox(0,0){ 2000}}%
\put(1452,723){\makebox(0,0){ 1800}}%
\put(1130,723){\makebox(0,0){ 1600}}%
\put(809,723){\makebox(0,0){ 1400}}%
\put(487,723){\makebox(0,0){ 1200}}%
\put(165,723){\makebox(0,0){ 1000}}%
\put(150,1266){\makebox(0,0)[r]{ 450}}%
\put(150,1202){\makebox(0,0)[r]{ 400}}%
\put(150,1138){\makebox(0,0)[r]{ 350}}%
\put(150,1074){\makebox(0,0)[r]{ 300}}%
\put(150,1009){\makebox(0,0)[r]{ 250}}%
\put(150,945){\makebox(0,0)[r]{ 200}}%
\put(150,881){\makebox(0,0)[r]{ 150}}%
\put(150,817){\makebox(0,0)[r]{ 100}}%
\put(150,753){\makebox(0,0)[r]{ 50}}%
\put(858,1893){\makebox(0,0)[r]{Hyams et al.}}%
\put(858,1971){\makebox(0,0)[r]{Kaminski et al.}}%
\put(1050,1419){\makebox(0,0){$\sqrt{s}\ \textrm{(MeV)}$}}%
\put(30,1765){%
\special{ps: gsave currentpoint currentpoint translate
270 rotate neg exch neg exch translate}%
\makebox(0,0)[b]{\shortstack{$\lvert S_{11} \rvert$}}%
\special{ps: currentpoint grestore moveto}%
}%
\put(1935,1479){\makebox(0,0){ 1600}}%
\put(1581,1479){\makebox(0,0){ 1400}}%
\put(1227,1479){\makebox(0,0){ 1200}}%
\put(873,1479){\makebox(0,0){ 1000}}%
\put(519,1479){\makebox(0,0){ 800}}%
\put(165,1479){\makebox(0,0){ 600}}%
\put(150,2022){\makebox(0,0)[r]{ 2}}%
\put(150,1894){\makebox(0,0)[r]{ 1.5}}%
\put(150,1766){\makebox(0,0)[r]{ 1}}%
\put(150,1637){\makebox(0,0)[r]{ 0.5}}%
\put(150,1509){\makebox(0,0)[r]{ 0}}%
\put(1178,2983){\makebox(0,0)[r]{BNL-E865}}%
\put(1178,3061){\makebox(0,0)[r]{NA48/2}}%
\put(1287,3253){\makebox(0,0){ 400}}%
\put(909,3253){\makebox(0,0){ 350}}%
\put(531,3253){\makebox(0,0){ 300}}%
\put(327,3223){\makebox(0,0)[r]{ 30}}%
\put(327,3119){\makebox(0,0)[r]{ 20}}%
\put(327,3016){\makebox(0,0)[r]{ 10}}%
\put(327,2912){\makebox(0,0)[r]{ 0}}%
\put(1652,2240){\makebox(0,0)[r]{Prom.}}%
\put(1652,2318){\makebox(0,0)[r]{Kaminski et al.}}%
\put(1652,2396){\makebox(0,0)[r]{Ke4}}%
\put(1050,2067){\makebox(0,0){$\sqrt{s}\ \textrm{(MeV)}$}}%
\put(30,2737){%
\special{ps: gsave currentpoint currentpoint translate
270 rotate neg exch neg exch translate}%
\makebox(0,0)[b]{\shortstack{$\delta_{11}\ \textrm{(deg)}$}}%
\special{ps: currentpoint grestore moveto}%
}%
\put(1801,2127){\makebox(0,0){ 1500}}%
\put(1132,2127){\makebox(0,0){ 1000}}%
\put(463,2127){\makebox(0,0){ 500}}%
\put(150,3318){\makebox(0,0)[r]{ 450}}%
\put(150,3203){\makebox(0,0)[r]{ 405}}%
\put(150,3087){\makebox(0,0)[r]{ 360}}%
\put(150,2972){\makebox(0,0)[r]{ 315}}%
\put(150,2857){\makebox(0,0)[r]{ 270}}%
\put(150,2742){\makebox(0,0)[r]{ 225}}%
\put(150,2626){\makebox(0,0)[r]{ 180}}%
\put(150,2511){\makebox(0,0)[r]{ 135}}%
\put(150,2396){\makebox(0,0)[r]{ 90}}%
\put(150,2281){\makebox(0,0)[r]{ 45}}%
\put(150,2165){\makebox(0,0)[r]{ 0}}%
\end{picture}%
\endgroup
 
\caption{Left panels, from up to down:  $\pi\pi\rightarrow \pi\pi$ phase shift, amplitude and the same 
for $\pi\pi\rightarrow K\bar{K}$. Right panels, from up to down: 
modulus squared of the S-matrix elements $\pi\pi\to\eta\eta'$ and $\pi\pi\to\eta\eta$. The last two figures 
correspond to the phase and modulus of the $K^-\pi^+\to K^-\pi^+$ scattering.\label{figura1}}
\end{figure}

With all these amplitudes, one can construct the $S$-matrix and calculate observables; in our case, these will be phase shifts
 and amplitude moduli.
 The curves resulting from our fit are depicted in Fig. \ref{figura1}.
  We use 13 parameters for about 373 experimental data, and a fair agreement with data is achieved.
   We have reduced the number of free parameters compared with other approaches in the literature 
   which do not employ (chiral) Lagrangians.

Once the observables are fitted, we can explore the $s-$complex plane to find the relevant poles of the
 amplitudes, and discuss their resonance\index{subject}{resonance} content.
  We present in Table \ref{table1} the masses and widths of the resonances that we find. 
  The agreement with the ones in the PDG is remarkable.

\begin{table}[h]
\caption{Parameters of resonances. On the left columns we have the masses and widths that we find. On the right ones, the values 
are  given by the PDG or the BES Collaboration.\label{table1}}
\begin{center}
\begin{tabular}{|c||c|c|} \hline
Resonance & Mass (MeV) & Width (MeV) \\ \hline\hline
$\sigma$ & 454 & 475 \\ \hline
$f_0(980)$ & 980 & 44 \\ \hline
$f_0(1370)$ & 1380 & 350 \\ \hline
$f_0(1500)$ & $\approx$ 1500 & $100-170$\\ \hline
$f_0(1710)$ & $\approx$ 1680 & $\approx$ 160 \\ \hline
$f_0(1790)$ & $\approx$ $1805^{\ }_{\ }$ & $\approx$ 390 \\ \hline
\end{tabular}
\begin{tabular}{|c|c|} \hline
Mass (MeV) & Width (MeV) \\ \hline\hline
400-1200 & 600-1200 \\ \hline
$980\pm 10$ & 40-100 \\ \hline
1200-1500 & 200-500 \\ \hline
$1507 \pm 5$ & $109 \pm 7$ \\ \hline
$1718 \pm 6$ & $137 \pm 8$ \\ \hline
$1790^{+40}_{-30}$ & $270^{+60}_{-30}$ \\ \hline
\end{tabular}
\end{center}
\end{table}

\end{document}